\documentclass[pdflatex,sn-mathphys-num]{sn-jnl}

\usepackage{xr}   
\usepackage{hyperref}

\usepackage{graphicx}%
\usepackage{multirow}%
\usepackage{amsmath,amssymb,amsfonts}%
\usepackage{amsthm}%
\usepackage{mathrsfs}%
\usepackage[title]{appendix}%
\usepackage{xcolor}%
\usepackage{textcomp}%
\usepackage{manyfoot}%
\usepackage{booktabs}%
\usepackage{algorithm2e}    
\SetKwProg{Init}{init}{}{}

\usepackage{algpseudocode}%
\usepackage{listings}%
\usepackage{booktabs}
\usepackage{siunitx}

\theoremstyle{thmstyleone}%
\theoremstyle{thmstyletwo}%

\theoremstyle{thmstylethree}%
\newtheorem{definition}{Definition}%

\begin{document}

\title[Article Title]{\hspace{2cm}Echoes of the hidden:\\ Uncovering coordination beyond network structure}


\author*[1,2]{\fnm{Shahar} \sur{Somin}}\email{shaharso@mit.edu}

\author[1]{\fnm{Tom} \sur{Cohen}}\email{tomco@mit.edu}
\author[1]{\fnm{Jeremy} \sur{Kepner}}\email{kepner@ll.mit.edu}
\author[1]{\fnm{Alex} \sur{Pentland}}\email{pentland@mit.edu}

\affil[1]{ \orgname{MIT}, \orgaddress{\city{Cambridge}, \state{MA}, \country{USA}}}

\affil[2]{ \orgname{Bar-Ilan University}, \orgaddress{\city{Ramat-Gan}, \country{Israel}}}

\abstract{

The study of connectivity and coordination has drawn increasing attention in recent decades due to their central role in driving markets, shaping societal dynamics, and influencing biological systems. 
Traditionally, \textit{observable} connections, such as phone calls, financial transactions, or social media connections, have been used to infer coordination and connectivity. 
However, incomplete, encrypted, or fragmented data, alongside the ubiquity of communication platforms and deliberate obfuscation, often leave many real-world connections \textit{hidden}. 
In this study, we demonstrate that coordinating individuals exhibit shared bursty activity patterns, enabling their detection even when \textit{observable} links between them are sparse or entirely absent.
We further propose a generative model based on the \textit{network of networks} formalism to account for the mechanisms driving this collaborative burstiness, attributing it to shock propagation across networks rather than isolated individual behavior.
Model simulations demonstrate that when \textit{observable} connection density is below 70\%, burstiness significantly improves coordination detection compared to state-of-the-art temporal and structural methods.
This work provides a new perspective on community and coordination dynamics, advancing both theoretical understanding and practical detection. 
By laying the foundation for identifying hidden connections beyond \textit{observable} network structures, it enables detection across different platforms, alongside enhancing system behavior understanding, informed decision-making, and risk mitigation.

}
\maketitle

\keywords{Coordination, community detection, network of networks, collaborative burstiness}

\section{Introduction}
Connectivity and coordination among entities have long been recognized as fundamental driving forces of critical processes across a wide array of domains.
In society, coordinated inauthentic behavior shapes public opinion through influence campaigns, rumors, and misinformation \cite{kumar2017army,shu2017fake,shu2019beyond,shu2020hierarchical}.
In finance, economic bubbles are sparked by coordinated actions \cite{sornette2023non,sornette2009stock,abreu2003bubbles}, while market stability and integrity are violated by collusive trading practices, such as money laundering schemes and coordinated market manipulations \cite{kasa2019improving, viswanath2015strength}.
Coordination among malicious actors during attack planning stages is vital for successful execution of terror \cite{campedelli2021learning,desmarais2013forecasting} and cyber \cite{sridhar2011cyber,pasqualetti2013attack,khraisat2019survey} attacks, making detection crucial for prevention and defense.
Even in biology, interactions among 
microorganisms have been shown to profoundly impact human health \cite{baishya2019selective,peters2012polymicrobial,baishya2021impact}.

Traditional content-based community and coordination detection methods \cite{alizadeh2020content,rheault2021efficient,im2020still, sahingoz2019machine, ashcroft2015detecting} face growing limitations due to the shift toward encrypted platforms, which restrict access to user-generated content and their private attributes.
The widespread use of generative AI further amplifies this challenge, by enabling effortless and instantaneous creation of numerous narrative variations, further reducing the effectiveness of content-based approaches. 
Given these challenges, interaction-based methods have gained renewed relevance for detecting coordination in content-restricted domains.
These include structure-based community detection employing \textit{observable} network connections \cite{blondel2008fast,raghavan2007near,girvan2002community} representation learning \cite{wang2018graphgan,grover2016node2vec,perozzi2014deepwalk,velickovic2019deep} and graph neural networks \cite{wang2021self,jia2019communitygan,michail2022detection,wei2024financial}, which leverage indirect connections and local network topology.
However, these methods struggle with fragmented or incomplete structural data and are further hindered by deliberate obfuscation, where malicious actors minimize both direct and indirect connections to evade detection.
These challenges intensify in cross-platform settings, where coordinating entities inherently lack observable connections, leaving key coordination patterns undetectable by structure-based methodologies.

In this study, we explore the dynamics of community members acting on different encrypted platforms.
Despite lacking explicit connections in this setting, we demonstrate that coordinating individuals present similar bursty dynamics.
Using two use cases, across encrypted financial and social platforms, we demonstrate that burstiness enables the effective detection of coordinating groups across platforms, outperforming state-of-the-art structural and temporal methods.
We further propose a generative model based on the  \textit{network of networks} framework \cite{kenett2014network,gao2014single, gao2011robustness} to capture the forces governing collaborative burstiness.
Our model suggests that distinct
domains are implicitly connected through coordinating individuals, enabling external events to propagate like shock waves across otherwise disconnected platforms. 
These shocks trigger timely responses from community members, resulting in collaborative burstiness that, as suggested by our robustness analysis, is inherently difficult to obscure without undermining coordination itself.
Model simulations reproduce empirical results, and show that unless coordination results in highly dense \textit{observable} network structure, it is significantly better detected by bustiness than by other structure-based community detection methods. 
By broadening the theoretical foundations of coordination and introducing a robust detection methodology, this work offers a new prism for understanding communities and coordination across encrypted, fragmented, and data-restricted environments, where traditional methods fall short.

\section{Results}

 \begin{wrapfigure}{R}{0.5\textwidth} 
    \centering
    \vspace{-0.65cm}
\includegraphics[width=0.5\textwidth]{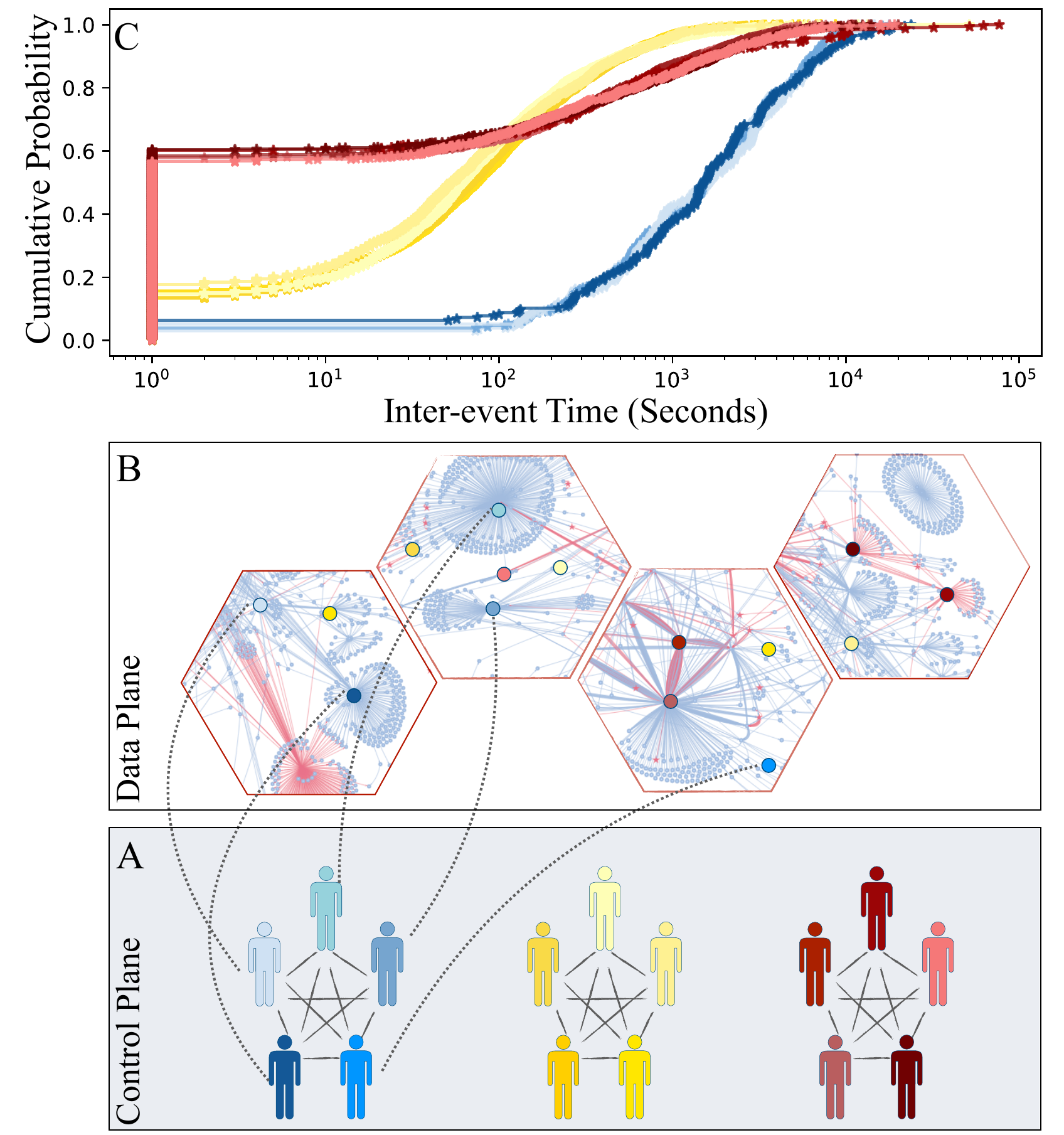}
\caption{\textbf{Community synchronization.} Presenting three layers of data. 
\textbf{a.} Control plane encompassing three communities, with members coordinating in the real world.
\textbf{b.} Data plane encompassing four distinct domains, with node color indicating community membership. Blue nodes illustrate the implicit connections across domains formed by the collaborating community members. 
\textbf{c.} Bursty dynamics of each community member, colors indicating community membership.
}\label{fig:fig1_motivation}
\vspace{-0.6cm}
\end{wrapfigure}

This study aims to better comprehend community synchronization and utilize it for detecting communities, even when explicit connections between their members are sparse or entirely absent. 
We pustulate that coordination among real-world community members participating in time-sensitive activities (e.g. financial transactions, social media posting) might be reflected in their temporal activity patterns, as members trigger each other into actions.
Given this premise, we analyze individual bursty patterns, manifested as the time difference between consecutive activities, and examine the associated inter-event time distribution for each individual (formally defined in Section \ref{sec:methods}). 
Fig. \ref{fig:fig1_motivation}a illustrates three real-world communities and their members' activity across four different domains (Fig. \ref{fig:fig1_motivation}b). 
Fig. \ref{fig:fig1_motivation}c demonstrates the synchronization in bursty dynamics among community members, and demonstrates how distinctive bursty dynamics enable differentiating between different groups.

\begin{wrapfigure}{R}{0.5\textwidth} 
    \centering
    \vspace{-0.6cm}
\includegraphics[width=0.5\textwidth]{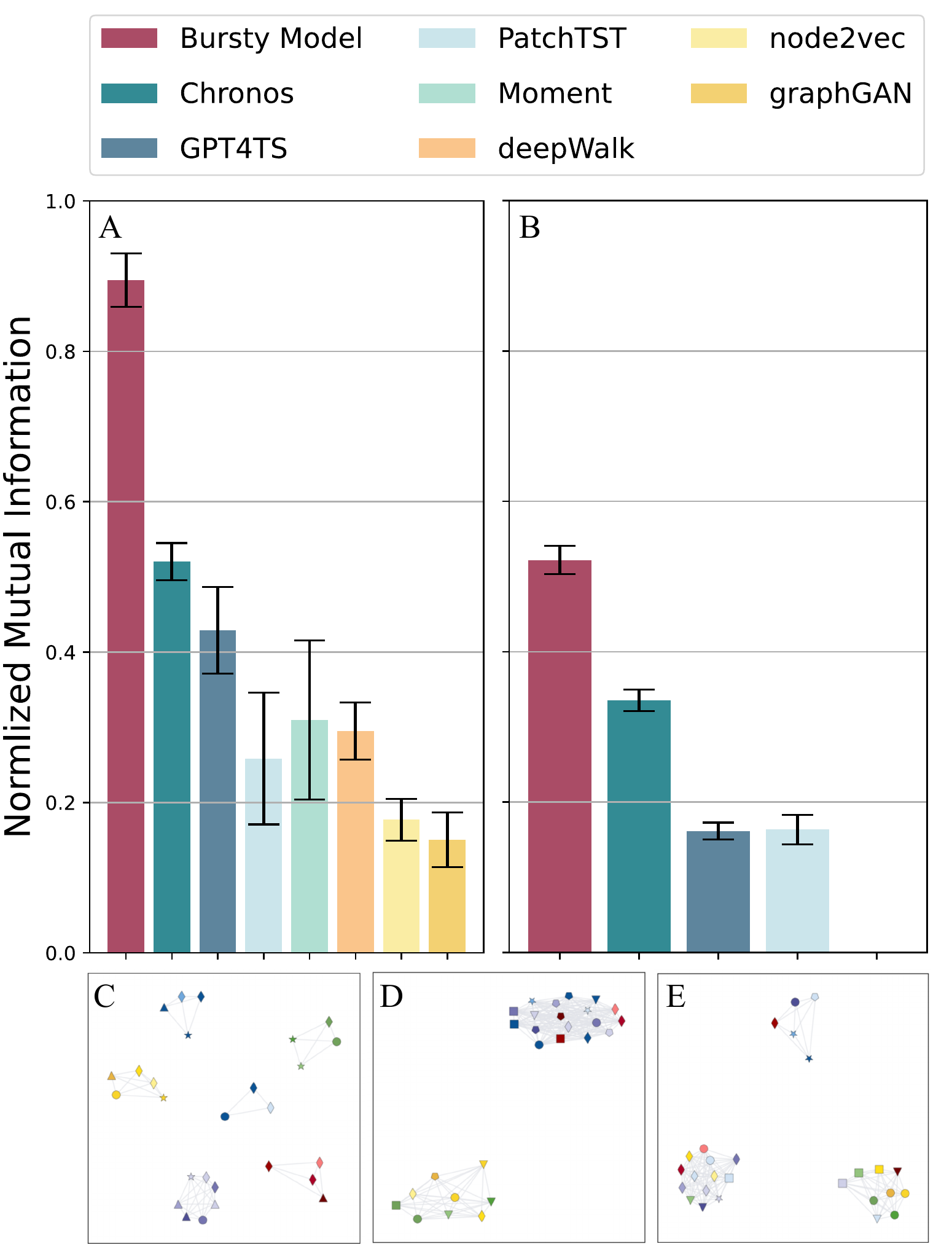}
\caption{\textbf{Performance evaluation.}
\textbf{a, b. } Normalized mutual information for community detection in the financial markets and in the social platforms experiments, correspondingly.
The bursty model outperforms all temporal (blue-shaded) and structural (yellow-shaded) models in both experiments. 
\textbf{c, d, e.} Example of cross-platform financial communities, detected using the bursty model, Chronos temporal model and deepWalk structural model, correspondingly. 
}\label{fig:fig2_perf}
\vspace{-0.5cm}
\end{wrapfigure}

We assess the similarity of users' bursty dynamics across domains (formally defined in Section \ref{sec:methods}) and evaluate its effectiveness in cross-domain community detection using two experimental settings.
First, we evaluate the community detection performance across different financial trading markets.
We run five weekly experiments, examining $7$ communities encompassing the weekly tradings of $61$ users, transacting on $49$ different financial trading markets, on top of the Ethereum blockchain \cite{buterin2014next, somin2022remaining,somin2022beyond}.
The second experimental setting encompasses community detection across social media platforms. 
In this setting, we examine $15$ communities, encompassing the weekly posts of $57$ users, on top of the Twitter, Telegram and Instagram platforms. 
Fig. \ref{fig:fig2_perf}a, b demonstrate that the inter-event bursty model detects cross-domain communities more effectively than structure-based \cite{grover2016node2vec,perozzi2014deepwalk,wang2018graphgan} and temporality-based \cite{goswami2024moment,ansari2024chronos,nie2023a,jin2024timellm} models both in the financial and the social experimental settings (consider further analysis in Fig. \ref{fig:multi_dom_blockchain_resolution}, supplementary information). 
The model's success in community detection on both experimental settings highlights its robustness and generalizability.
In order to assess the effectiveness of burstiness in community detection when structural data and explicit links are available, we compare the inter-event bursty model to temporal and structural baselines in a single-domain setting (consider Fig. S\ref{fig:single_dom_blockchain} in supplementary materials). 
We show that in this setting as well, the bursty model outperforms all baseline models, demonstrating its ability to detect coordination even when explicit structural connections between community members are available.

In light of the empirical results showing that key coordination patterns remain undetectable by structure- and temporality-based methods, we aim to understand the underlying mechanisms driving coordination, enabling it to persist beyond explicit structures while manifesting in bursty dynamics.
Building on the concept of \textit{network of networks} \cite{kenett2014network,gao2014single, gao2011robustness}, we propose that distinct
domains are implicitly connected through coordinating profiles or individuals (consider motivating Fig. \ref{fig:fig1_motivation}a,b).
These hidden connections enable external events to propagate like shock waves, influencing actions across otherwise disconnected platforms. 
We propose a new generative model to describe the network formation process, while considering both the underlying real-world communities alongside the timing of individual activity. 
Our model extends the degree-corrected Stochastic Blockmodel \cite{karrer2011stochastic} by incorporating temporality as a factor influencing node activation.
Specifically, at each iteration a source node is selected based on a probability that depends on the elapsed time since the node's last activity and the last activity of the other members in its community, representing how community members trigger each other into action.  
The target node is selected as in \cite{karrer2011stochastic}, where the parameter $\lambda\in[0,1]$ interpolates linearly between choosing the destination node from the source's community, or randomly from the entire node-set. 
In both cases, the target is chosen according to the preferential attachment mechanism \cite{barabasi1999emergence}. 

\begin{figure}[h]
\centering
\vspace{-5pt}
\includegraphics[width=0.85
\textwidth]{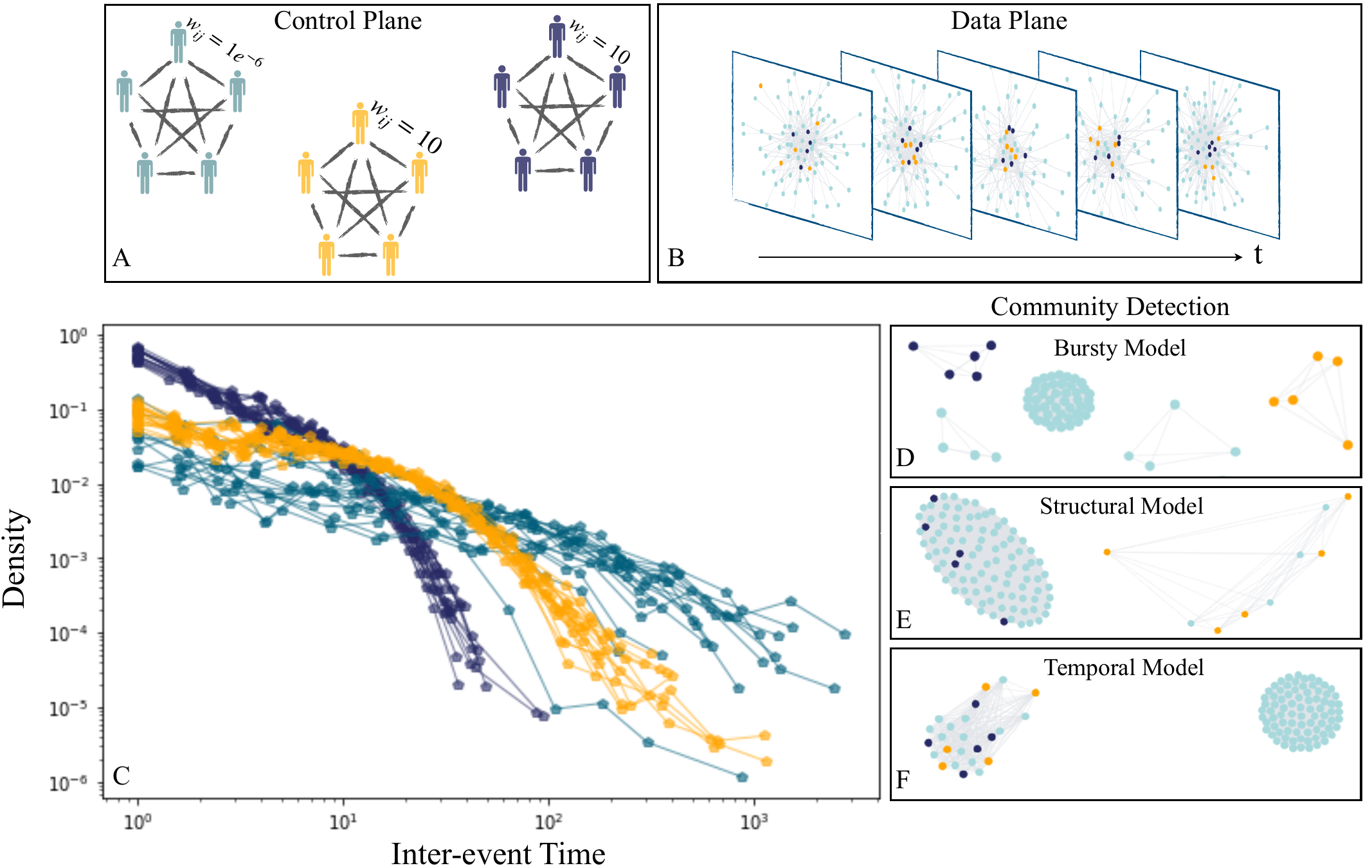}
\caption{\textbf{Generative model simulations.}
        \textbf{a.} Simulation control plane, encompassing three communities of varying weights.
        \textbf{b.} Temporal networks in the data plane, resulting from bursty-SBM model simulations.    
        \textbf{c.} Inter-event distributions of individual community members, colors indicating community membership.
        \textbf{d, e, f.}
        Simulation communities detection by the bursty, structural and temporal models, correspondingly.}\label{fig:fig3_simulations}
\end{figure}

We generate synthetic networks using the suggested model, with three different communities in the control plane, illustrated in \ref{fig:fig3_simulations}a, b. 
Encouragingly, the model reproduces the bursty collective dynamics of the different nodes, effectively differentiating between the communities, as demonstrated in Fig. \ref{fig:fig3_simulations}c. 
Fig. \ref{fig:fig3_simulations}d,e and f present the detected communities by the bursty model, a baseline structural model and a baseline temporal model, respectively, demonstrating the bursty model's significantly improved community detection ability. 
Consider comparisons against various baseline models in Table \ref{tab:simulation_sota_nmi}, supplementary materials.

\begin{wrapfigure}{R}{0.45\textwidth} 
    \centering
    \vspace{-0.0cm}
\includegraphics[width=0.45\textwidth]{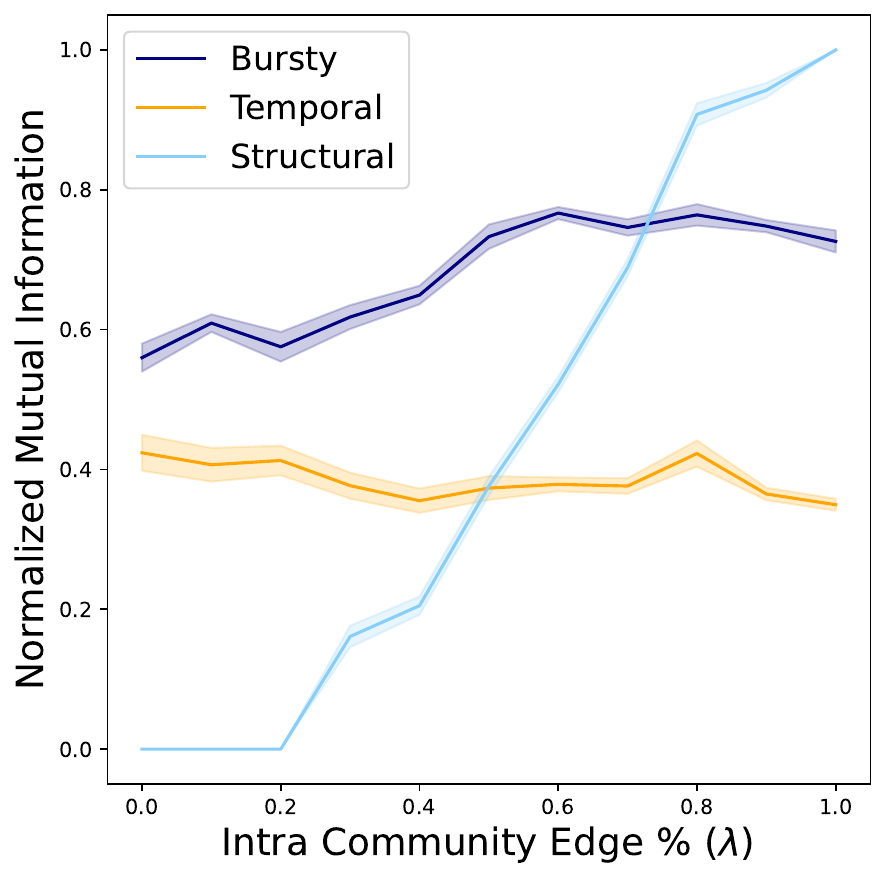}
\caption{\textbf{Density effect on community detection}
Community detection performance comparison of the bursty model (purple curve) to temporal (orange curve) and structural (light blue curve) models, and their dependence on intra-community edge percentage. 
The bursty model significantly outperforms baselines when less than $70\%$ of edges in the data plane are between community members. 
}\label{fig:fig4_sim_nmi_vs_lambda}
\vspace{-0.5cm}
\end{wrapfigure}
This modeling also enables characterizing the limits of structure-based community detection. 
Specifically, we examine the dependence of community detection performance on the density of communities in the
data plane.
Fig. \ref{fig:fig4_sim_nmi_vs_lambda} depicts the Normalized Mutual Information (NMI) of the bursty model, compared to structural and temporal baselines, as a function of $\lambda$, the parameter controlling intra-community edge percentage.
This analysis demonstrates that when real-world communities manifest as highly dense in the data plane ($\lambda \geq 0.7$), structure-based models outperform both the bursty model and other temporal baselines. 
However, as community density decreases ($\lambda < 0.7$), the bursty model detects them significantly better than structure-based methods. 
This underscores its ability to reveal latent connections that remain undetectable through explicit structural links, enhancing our capacity to identify coordination in complex systems.

\section{Discussion}

Our study sheds light on the origins of coordination manifestation in temporal patterns
framed through the \textit{network of networks} \cite{kenett2014network,gao2014single,
gao2011robustness} formalism. 
Our model suggests that coordinating individuals act as bridges between seemingly disconnected domains, forming implicit connections that enable events to propagate as shock waves throughout the system and  trigger actions across domains.
This framework, accompanied by a \textit{forget} mechanism, offers a possible explanation to the emergence of \textit{collaborative burstiness}, where similar bursty patterns arise among coordinating individuals, even when their activities are not simultaneous.
Previous studies have modeled bursty human activity primarily relying on the level of isolated individuals. 
Common approaches include priority-queuing processes \cite{barabasi2005origin,vazquez2006modeling,gabrielli2007invasion}, Poisson dynamics \cite{malmgren2009universality,malmgren2008poissonian}, and preferential linking \cite{gonccalves2008human}. However, these models largely overlook the role of environmental influences.
By incorporating interactions that trigger cascades of activity through a network of networks, our approach provides a broader perspective on the origins of bursty behavior, capturing the notion of collaborative burstiness.

A natural question arises as to whether coordination signals emerging from bursty temporal dynamics can be effectively concealed.
In online communication platforms, where timely responses are integral, masking coordination is inherently limited, as individuals tend to react promptly to external shocks.
While adversaries may attempt to obscure coordination by distributing actions over time or employing hidden agents, such strategies are often impractical. For instance, delayed financial transactions may trigger suspicion in anti-money laundering frameworks, and temporally dispersed actions can diminish the operational impact of coordinated attacks. These considerations suggest that efforts to conceal coordination may ultimately compromise the effectiveness of the coordinated activity itself.
Consider the supporting robustness analysis performed in our previous paper \cite{somin2025identity}, demonstrating that even upon substantial activity omission, users' burstiness similarity is maintained.

Our empirical study further distinguishes between structural and temporal manifestations of coordination. 
In the cross-domain setting, temporal models significantly outperform structural ones in detecting coordination (Fig. \ref{fig:fig2_perf}), whereas in the single-domain setting, structural models perform better (Fig. \ref{fig:single_dom_blockchain}, supplementary materials). 
While the bursty model outperforms all methods in both cases, the contrast between structural and temporal performance across single- and multi-domain scenarios is of particularly interest. 
We postulate that in a single domain, structural models benefit from neighborhood similarity and explicit connections between coordinating individuals, which enhance their effectiveness.
This hypothesis is further supported by our generative model, which demonstrates the significant impact of structural density on coordination detection performance (Fig. \ref{fig:fig4_sim_nmi_vs_lambda}).

This study models the role of latent connectivity in the control plane on coordination dynamics observed in the data plane, raising several open questions for future research. 
An important direction involves refining the theoretical framework of coordination by developing more expressive generative models, such as mixed-membership block models, to better capture scenarios where individuals participate in multiple communities.
Furthermore, the broader implications of latent connectivity on network resilience, centrality, and influence are of particular interest. 
Investigating their role in shaping robustness under failures and attacks, identifying influential nodes beyond explicit connections, and improving models of information spread could provide deeper insights into network evolution and coordination processes.

In addition to these theoretical advancements, this study also has important practical implications. 
Developing privacy-preserving methods for identifying latent connections and influences across domains could help mitigate financial risks, counter external societal influences, and enhance the detection of cross-domain effects, where activity in one domain triggers coordinated responses in another, such as financial instability driven by misinformation spreading on social media. 
The proposed coordination detection methodology also holds promise for cyber defense and risk mitigation. 
By enabling the early identification of adversarial activities, both within and across encrypted domains, these findings could support a proactive cybersecurity approach, strengthening threat prevention before system breaches occur.
More broadly, this work underscores the need to move beyond traditional structural analysis toward dynamic, behavior-based approaches, offering new avenues for understanding and mitigating hidden coordination in complex systems.

\paragraph{Funding Acknowledgment}
Research was sponsored by the United States Air Force Research Laboratory and the Department of the Air Force Artificial Intelligence Accelerator and was accomplished under Cooperative Agreement Number FA8750- 19-2-1000. The views and conclusions contained in this document are those of the authors and should not be interpreted as representing the official policies, either expressed or implied, of the Department of the Air Force or the U.S. Government. The U.S. Government is authorized to reproduce and distribute reprints for Government purposes notwithstanding any copyright notation herein.

\newpage
\section{Methods}\label{sec:methods}
\subsection{Datasets}
\subsubsection*{Financial markets data}
We consider the Ethereum blockchain \cite{buterin2014next,somin2022remaining,somin2022beyond,somin2020network}
as our financial dataset. 
This encrypted financial ecosystem enables the trading of tens of thousands of different crypto-tokens, using an Ethereum wallet. Broadly, a crypto wallet is a digital
tool that securely stores and manages the user’s cryptocurrency holdings, allowing the
user to send, receive, and monitor their digital assets on blockchain networks.
To establish ground truth for coordinating groups on the Ethereum blockchain, we use data from a blockchain explorer \cite{etherscan}, which serves as a public ledger interface, to retrieve metadata on blockchain wallets. 
We tag the wallets based on their associated organization or service provider, forming groups that, by definition, engage in coordinated activity as part of the same entity.
We consider two different experimental settings over this dataset. 
\paragraph{Single domain setting:} Considering $5$ days of trading activity on top of the entire Ethereum blockhcain, encompassing the activity of $9$ daily coordinated groups consisting of a total of $58$ traders (see details in Table \ref{si:single_dom_comms}, supplementary materials).
This setting contain temporal data on an individual level granularity and network data, where an edge $(u,v)\in V$ represents that user $u$ sold any crypto-token to user $v$.  


\paragraph{Multi domain setting:} Considering $5$ weeks of trading activity of $7$ weekly coordinated groups, encompassing $61$ traders across $49$ financial trading domains (see details in Table \ref{si:multi_dom_comms}, supplementary materials).
We refer to a \textit{financial trading market} $D^i_{\tau}$ as encompassing 
all of the trading activity related to the respective crypto-token $c_i$ during time period $[\tau, \tau+\Delta \tau ]$ where $\Delta \tau$ stands for single day length:
\begin{equation}
        D^i_{\tau} = \{u : u \text{ bought or sold } c_i \text{ in }[\tau, \tau+\Delta \tau ] \}
\end{equation}
This setting contain temporal data on an individual level granularity and network data, where an edge $(u,v)\in V_i$ represents that user $u$ sold crypto-token $i$ to user $v$. 


%

\subsubsection*{Social platforms data}
The dataset contains activity metadata of profiles from across Twitter (X), Telegram, and Instagram, collected over two months (June 8, 2024 - July 20, 2024). 

\paragraph{Single domain setting:} 
We examine four Telegram groups, encompassing the posting activity of $14415$ profiles.
We label a group of profiles as belonging to the same community if they post in the same Telegram group (see details in Table \ref{si:single_dom_comms_telegram}, supplementary materials). 
The experiment on this dataset entailed analyzing six weekly activity snapshots of these users, where each snapshot contains merely temporal data in the form of individual posting times, and lacks network data in the form of re-posts, likes and other connections.

\paragraph{Multi domain setting:} 
We examine $15$ coordinated groups, encompassing $33$ Twitter profiles, $20$ Telegram profiles, and $4$ Instagram profiles (see details in Table \ref{si:SM_cd_comm}, supplementary materials). 
We label a group of profiles as belonging to the same community if their profile names, biographies, or other publicly available information (articles, Linktree) clearly indicate they belong to, represent, or work under the same organization.
The first stage of the community-tagging process was surfacing pairs of profiles with very similar names and affirming their relationship manually by reviewing their profiles.
Afterwards, we expanded the community with affiliates or subsidiaries of the profiles, which we found via biographical information, direct mentions, or in-platform suggestions.
The experiment on this dataset entailed analyzing six weekly activity snapshots of these users on top of both social platforms. 
Each snapshot contains merely temporal data in the form of individual posting times, and lacks network data in the form of retweets, re-posts, likes and other connections.



\subsection{Burstiness based coordination detection}
We propose exploiting individual temporal data for detecting coordinating individuals across different domains.
Specifically, we analyze individual bursty patterns, manifested as the time difference between any two consecutive activities of each profile. Formally:
\begin{definition}\label{def:inter-event}
    Given a time period $[\tau, \tau+\Delta \tau]$ and a profile $u_d$ in domain $D$ we denote 
    the sequence of their activity times  $A^{u_d}_{\tau}\subset \left[\tau,\tau+\Delta \tau\right]$ by:
    \begin{equation}
        A^{u_d}_{\tau} = (t_0^{u_d},...,t_m^{u_d})
    \end{equation}
    An inter-event time period is defined as the time difference between two consecutive activities of $u_d$:
    \begin{equation}
        \Delta t_i^{u_d} = t_{i}^{u_d} - t_{i-1}^{u_d}
    \end{equation}
    The inter-event time sequence is defined by:
    \begin{equation}
        S^{u_d}_{\tau}=(\Delta t_1^{u_d},...\Delta t_m^{u_d})
    \end{equation}
    The cumulative distribution function of the inter-event sequence is defined as:
    \begin{equation}
        Q^{u_d}_{\tau}(\Delta t) = \frac{|\delta \in S^{u_d}_{\tau}: \delta \leq \Delta t|}{m}
    \end{equation} 
\end{definition}

\newpage
The similarity between the established inter-event time distributions of any two profiles is estimated by the Kolmogorov-Smirnov (KS) statistic:  

\begin{definition}\label{def:bursty-model}
Let $u_{d_1}\in D_1^{\tau}$ and $v_{d_2} \in D_2^{\tau}$, and their corresponding inter-event time distributions $Q^{u_{d_1}}_{\tau}$ and $Q^{v_{d_2}}_{\tau}$. 
The KS-statistic is defined as the maximal difference between their distributions:
\begin{equation}
    KS_{\tau}(u_{d_1},v_{d_2})= \sup_{\Delta t} |Q^{u_{d_1}}_{\tau}(\Delta t)-Q^{v_{d_2}}_{\tau}(\Delta t)|
\end{equation}
\end{definition}

Next, we build a weighted \textit{similarity network}, with weights based on the established scores representing the similarity between individual burstiness. 
We apply the Louvain community detection algorithm \cite{blondel2008fast} in order to detect groups of highest similarity. 
We evaluate the coordination detection performance by quantitatively comparing the communities found by the examined algorithm and the ground truth communities. 
We use the Normalized Mutual Information to assess:
\begin{equation}
    NMI(X,Y) = \frac{2I(X;Y)}{H(X)+H(Y)}
\end{equation}
where $X$ is the estimated partition of individuals into groups, $Y$ is their ground truth partition, $I(X;Y)$ is the mutual information and $H(X), \, H(Y)$ are the associated entropy.

\subsection{Bursty-Corrected SBM Generative Model}
We propose a new generative model to describe the network formation process, accounting for temporality as well as community membership as a factor controlling node activation. 
Community membership in the control plane is represented by a blockmodel $A_w$, where each community is represented by a weighted block, whose weight describes the strength of connections between the community members. 
At each iteration of the model, source and target nodes are selected. 
The probability to choose a source node depends on the time that has passed since the node's last activity, held by a vector $z_t$ and the last activity of his control plane neighbors, represented by the underlying SBM. 
The target node is selected similarly to the form originally presented in DG-SBM \cite{karrer2011stochastic}, where $\lambda \in [0,1]$ linearly interpolates between choosing the target node from the source’s community (block), or randomly from the entire node-set. 
The target is chosen from the subset of potential nodes according to the preferential attachment mechanism \cite{barabasi1999emergence}.

\begin{algorithm}[H]
\DontPrintSemicolon
\SetAlgoLined
\SetKwInOut{Input}{input}\SetKwInOut{Output}{output}
\Input{$A_w$, $A_{rand}$, $z_t \gets \text{Random}(0,1000)^{nT} $, $d \gets \vec{0}$}
\Output{$\{G_t(V_t,E_t)\}_{t\in[T]}$}
\For{$t\in[T]$}{
    \For{$i\in[n]$}{
        $P(v_{source})\propto A_w \cdot z$\;
        $\lambda \leftarrow \text{Random}(0,1)$\;
        $P(v_{target})\propto \lambda A_w \cdot d + (1-\lambda)A_{rand}\cdot d$\;
        $V_t \leftarrow V_t \cup \{v_{source}, v_{target}\}$\;
        $E_t \leftarrow E_t \cup \{(v_{source}, v_{target})\}$\;
        $d[v_{target}] \leftarrow d[v_{target}]+1$\;
        $d[v_{source}] \leftarrow d[v_{source}]+1$\;
        \For{$u \in \cup_{\tau<t} V_{\tau} $}{
            $z[u]\leftarrow z[u]+1$
        }
        $z[v_{source}]\leftarrow 0$\;
        
    }
}
\caption{Bursty Corrected SBM}
\end{algorithm}

\subsection{Comparison to Baseline Models}
We compare the coordination detection performance of the bursty model to various state-of-the-art models.

\subsubsection{Structure-based Models}
For the coordination detection within a single platform setting, we compare against four baseline models, which rely on network data. 
\begin{enumerate}
    \item \textbf{Louvain\cite{blondel2008fast}} A hierarchical community detection method optimizing modularity by iteratively grouping nodes into communities. It first, locally optimizes modularity by merging nodes into communities, then it aggregates these communities, repeating the process until modularity gain is maximized.
    \item \textbf{Label Propagation Algorithm (LPA)\cite{raghavan2007near}} An unsupervised community detection method that iteratively assigns labels to nodes based on the majority label of their neighbors. Through repeated updates, densely connected nodes converge to the same label, assisting to identify community structures.
\item\textbf{deepWalk\cite{perozzi2014deepwalk}} A node embedding method that learns low-dimensional representations of graph nodes by considering random walks on the graph as sentences and applying a skip-gram model to capture structural similarities. This approach enables nodes with similar connectivity patterns to have closer embeddings. We apply cosine similarity on all embedding pairs to form scores, indicating the certainty that both profiles are coordinating with each other.
Next, we build a weighted \textit{similarity network}, with weights based on the established deepWalk similarity scores. 
We apply the Louvain community detection algorithm in order to detect groups of highest similarity. 
\item\textbf{node2vec}\cite{grover2016node2vec}
   A node embedding method extending DeepWalk by introducing a biased random walk strategy, allowing it to explore local (depth-first) and global (breadth-first) neighborhoods. We apply cosine similarity on all embedding pairs to form scores, indicating the certainty that both profiles are coordinating with each other.
   Next, we build a weighted \textit{similarity network}, with weights based on the established node2vec similarity scores. 
    We apply the Louvain community detection algorithm  in order to detect groups of highest similarity. 
\item\textbf{graphGAN}\cite{wang2018graphgan} a generative adversarial framework for graph representation learning, where a generator models the underlying connectivity distribution of nodes and a discriminator distinguishes between real and generated edges. Through adversarial training, GraphGAN learns node embeddings capturing local and global graph structures.    
We apply cosine similarity on all embedding pairs to form scores, indicating the certainty that both profiles are coordinating with each other.
Next, we build a weighted \textit{similarity network}, with weights based on the established graphGAN similarity scores. 
We apply the Louvain community detection algorithm  in order to detect groups of highest similarity. 
   \end{enumerate}

For the cross-domain coordination detection setting we compare against deepWalk and node2vec and graphGAN which do not rely on explicit network connections to deduce coordination, but utilized similarities between node structural characteristics, allowing to generate node representation despite acting across different domains.

\subsubsection{Temporal Models}

Foundation models are large-scale, pre-trained machine learning models, typically based on transformer architecture \cite{raffel2020exploring,vaswani2017attention}, designed for a wide range of downstream tasks across diverse domains.
Building upon the success of foundation models in language and vision, foundation models for time series are large pre-trained models that capture complex patterns in temporal data across diverse domains and used for tasks like forecasting, sequence classification, anomaly detection, and imputation. 
In recent years, numerous foundation models for time-series analysis have been introduced \cite{liang2024foundation}, each employing unique methodologies to enhance forecasting accuracy and efficiency. 

In this paper, we benchmark our bursty model against four state-of-the-art foundation models for time series analysis: CHRONOS, MOMENT, PatchTST, and GPT4TS. 
These models have been trained on datasets from various domains, spanning electricity, traffic, weather, and health, as well as the UCL and UEA collections.
Given the scarcity of labeled data in many domains in time-series analysis, a particularly critical challenge for foundation models is zero-shot learning, which refers to the ability to perform tasks on unseen data without requiring additional training or fine-tuning \cite{larochelle2008zero}. Benchmarking on zero-shot tasks evaluates the robustness and versatility of pre-trained models, highlighting their potential for real-world applications where task-specific data is limited or unavailable. 
Importantly, all models were reported to support zero-shot learning, and were evaluated on such settings.

\begin{enumerate}
    \item \textbf{GPT4TS \cite{zhou2023one}} 
    Leverages pre-trained language models, specifically GPT-2, for time-series analysis. 
    By treating time-series data similarly to textual data, GPT4TS applies the strengths of language models for forecasting tasks. The model fine-tunes the embedding layer, normalization layers, and output layer of GPT-2 to accommodate time series inputs, while keeping the self-attention and feedforward layers of the model unchanged. 
    The model’s short- and long-term forecasting performance has been tested on ETT, Weather, ILI, and ECL datasets \cite{sun2024test} (Zhou et al, 2023; Ma et al, 2024), including zero-shot experiments on the ETT-H and -M datasets.

\item \textbf{PatchTST \cite{nie2023a}} A prominent transformer-based model tailored for long-term, multivariate time series forecasting. 
A key feature of the model is segmenting time-series data into subseries-level patches, which serve as input tokens to the transformer. 
This reduces the computational load by decreasing the length of input sequences and thereby lowering the time and space complexity associated with self-attention mechanisms. 
For zero-shot performance evaluation, it is tested on a diverse array of datasets spanning healthcare, finance and economics, retail and energy.

\item \textbf{Chronos \cite{ansari2024chronos}} A family of pre-trained probabilistic time-series models that adapt language model architectures for time-series forecasting and encoding tasks, with parameter sizes ranging from $8$ million (tiny) to $710$ million (large). 
The model tokenizes time series data through scaling and quantization, enabling the application of language models to the data. 
The models are pre-trained on a collection of $13$ datasets, encompassing energy, transportation, weather, and web traffic and have been evaluated on a wide range of datasets. 
For zero-shot performance, it is further tested on a wider array of datasets, including healthcare, retail, banking, and more.

\item \textbf{Moment \cite{goswami2024moment}} A family of foundation models designed for general-purpose time-series analysis. 
It addresses challenges in pre-training large models on time-series data by compiling a diverse collection of public datasets, termed the \textit{Time Series Pile}, which spans $5$ large public databases. 
Moment employs multi-dataset pre-training to capture diverse time-series characteristics, enhancing its adaptability across various tasks such as forecasting, classification, anomaly detection, and imputation. 

\end{enumerate}

After extracting and grouping temporal information per user per time period, we compute embedding similarity between every pair of cross-platform users in the following manner: First, we align the temporal sequences with the desired model input length. This process includes left-padding or truncating time steps to match the desired model input (typically, $N = 512$). 
We then pass the padded tensor, adjusting dimensions as per model requirements, into the selected model in batches of either $64$ (CHRONOS, MOMENT, PatchTST) or $16$ (GPT4TS) to accommodate GPU constraints. 
If applicable, we apply mean pooling to the output embeddings to achieve a one-dimensional vector representation per user. 
Computing the similarity is then straightforward: we apply cosine similarity between each pair of users, skipping self- and same-platform comparisons (as we are only interested in cross-platform identity matching) to derive a final similarity score.
Next, we build a weighted \textit{similarity network}, with weights based on the established scores by each model. We apply the Louvain community detection algorithm \cite{blondel2008fast} in order to detect groups of highest similarity.

\subsection{Data Availability}
The datasets generated during and analyzed during the current study are available from the corresponding author on reasonable request. 

\paragraph{Funding Acknowledgment}
Research was sponsored by the United States Air Force Research Laboratory and the Department of the Air Force Artificial Intelligence Accelerator and was accomplished under Cooperative Agreement Number FA8750- 19-2-1000. The views and conclusions contained in this document are those of the authors and should not be interpreted as representing the official policies, either expressed or implied, of the Department of the Air Force or the U.S. Government. The U.S. Government is authorized to reproduce and distribute reprints for Government purposes notwithstanding any copyright notation herein.

\newpage
\section*{Supplementary Information}

\section*{Coordination detection within a single domain}
We examine the ability to detect coordinating groups acting on a single domain. 
To this end, we performed experiments within a financial platform and within a social platform.
\begin{enumerate}
    \item \textbf{Coordination within a financial platform:} We perform five experiments on daily financial data. 
    The ground truth consists of $9$ communities, encompassing the trading activity of $58$ individual accounts. 
    \item \textbf{Coordination within a social platform:} We perform eight experiments on weekly Telegram data. The ground truth consists of $4$ non-overlapping communities, encompassing the activity of $14415$ Telegram profiles.  We label a group of profiles as belonging to the same community if they post in the same Telegram group.
\end{enumerate}

\begin{figure}[h]
\centering
\vspace{-5pt}
\includegraphics[width=0.75
\textwidth]{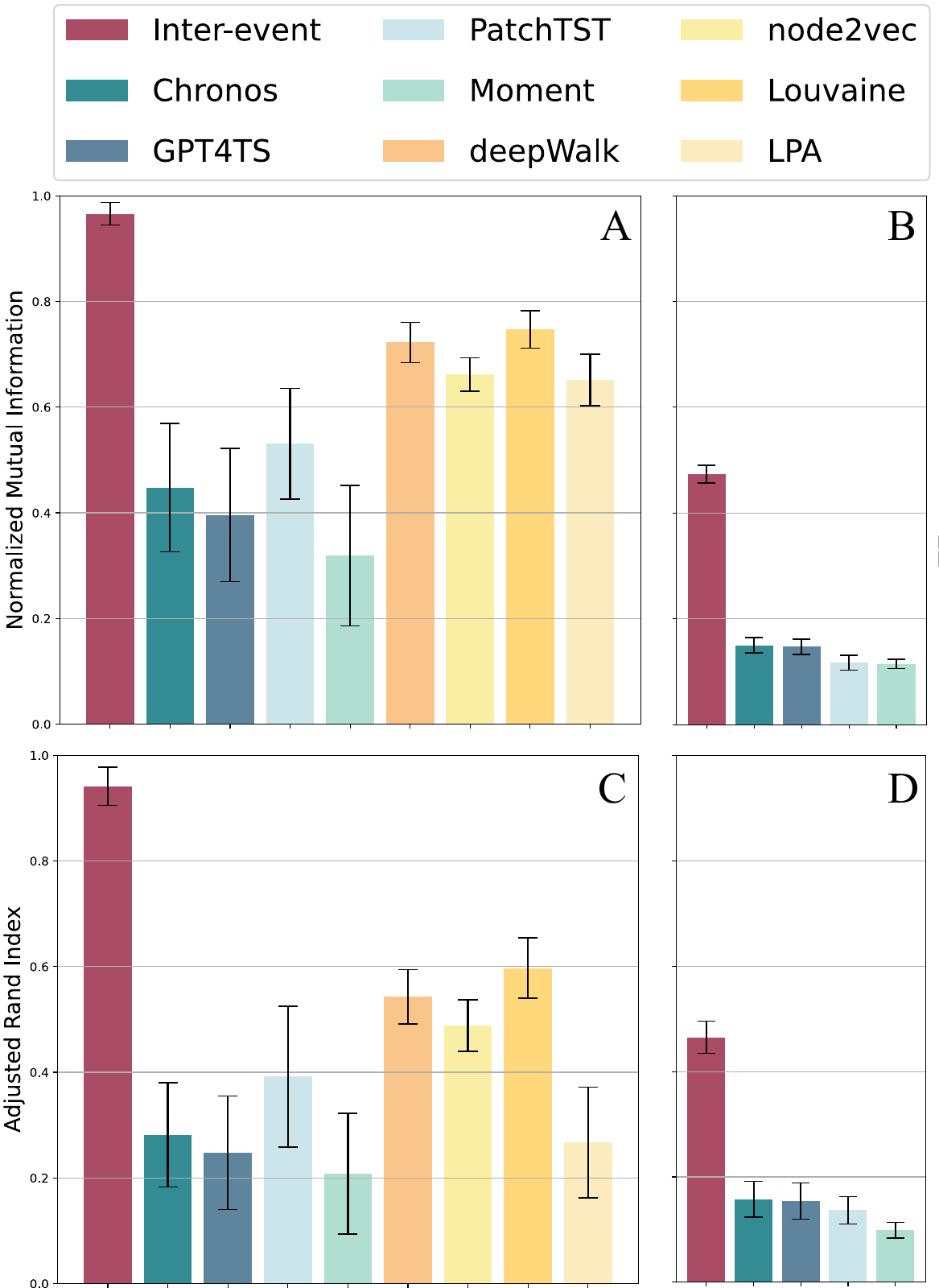}
\caption{\textbf{Coordination detection in a single financial market}
Comparing the performance of the inter-event bursty model to temporal (blue-shaded) and structural (yellow-shaded) state of the art models.
        \textbf{a.} Normalized Mutual Information (NMI) for coordination detection.
        \textbf{b.} Adjusted Rand Index (ARI) for coordination detection. 
        The bursty model outperforms all baseline models.  
}\label{fig:single_dom_blockchain}
        \vspace{-0.1cm}
\end{figure}

Fig. \ref{fig:single_dom_blockchain} compares the coordination detection performance of the inter-event bursty model to structural \cite{perozzi2014deepwalk,grover2016node2vec,blondel2008fast,raghavan2007near} and temporal \cite{zhou2023one,nie2023a,ansari2024chronos,goswami2024moment} baselines, for both experiment types.
This analysis demonstrates the effectiveness of accounting for individual bursty activity patterns for community detection, even in a setting where structural data and explicit links between individuals exist. 

\section*{Coordination detection across domains}
\begin{figure}[h!]
\centering
\vspace{-5pt}
\includegraphics[width=0.95
\textwidth]{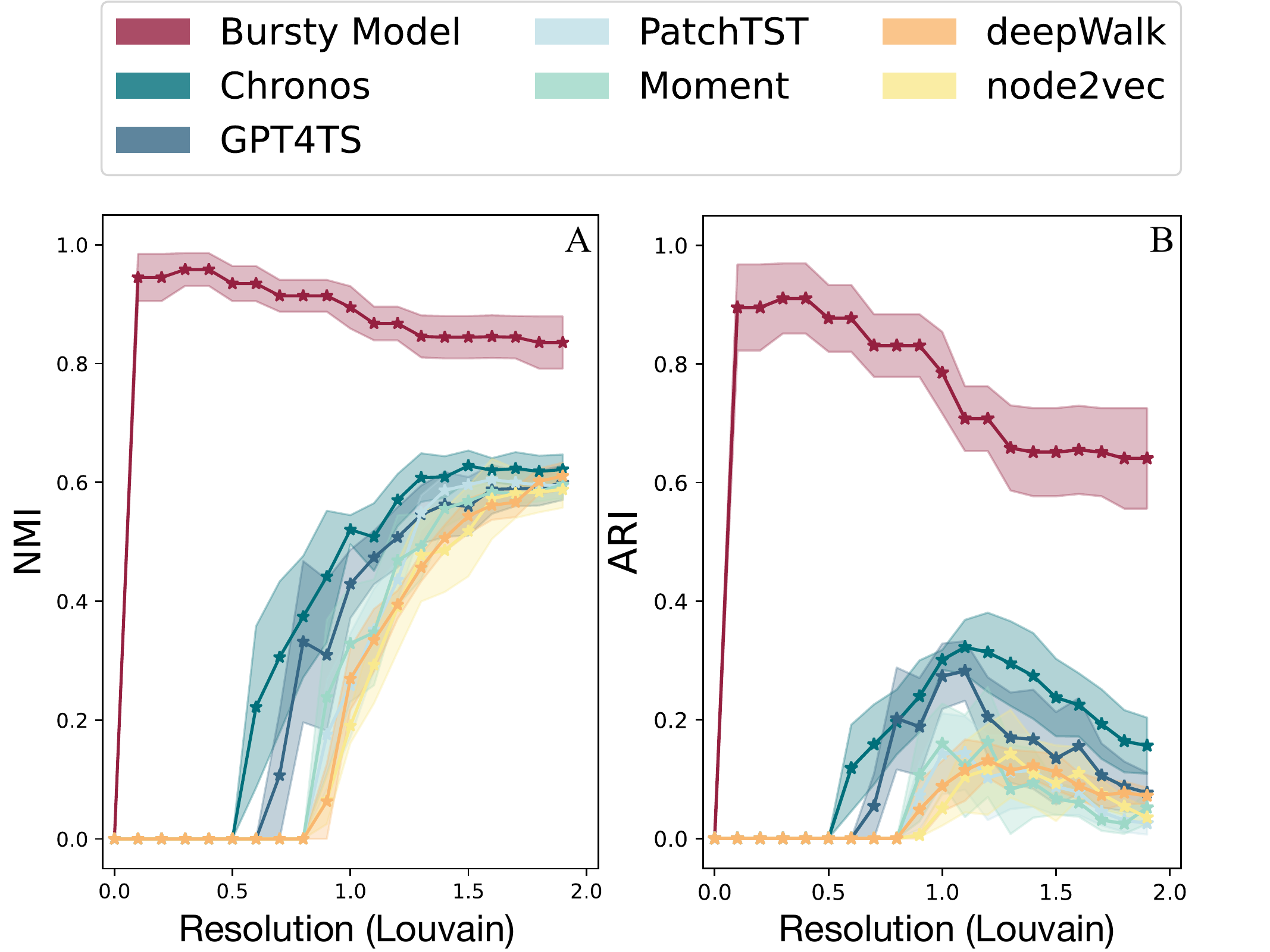}
\caption{\textbf{Coordination detection across multiple financial market, dependence on Louvain resolution}
Comparing the performance of the inter-event bursty model to temporal (blue-shaded) and structural (yellow-shaded) state-of-the-art-models, presenting the dependence on the resolution chosen for the Louvain community detection on the similarity networks.
        \textbf{a.} Normalized Mutual Information (NMI) for coordination detection as a function of the resolution.
        \textbf{b.} Adjusted Rand Index (ARI) for coordination detection as a function of the resolution. 
        The bursty model outperforms all baseline models, in all resolution levels.  }\label{fig:multi_dom_blockchain_resolution}
        \vspace{-0.4cm}
\end{figure}

In the cross-domain setting we use the Louvain clustering algorithm on top of similarity networks we construct, based on burstiness similarity, as well as based on temporal and structural similarities.
The main analysis was established using a Louvain community detection algorithm \cite{blondel2008fast}, with a resolution parameter $r=1$. 
We wish to affirm the performance of our methodology by assessing the dependence of coordination detection performance on the chosen resolution parameter.
Panels A and B in Fig. \ref{fig:multi_dom_blockchain_resolution} present the dependence of the NMI and ARI correspondingly on the Louvain resolution for each of the examined algorithms.
This analysis demonstrates that inter-event burstiness detects the coordinating groups significantly better than the examined baseline models, across the entire range of examined resolution values.

\section*{Bursty-Corrected Stochastic Blockmodel}
\begin{figure}[h]
\centering
\vspace{-5pt}
\includegraphics[width=0.95
\textwidth]{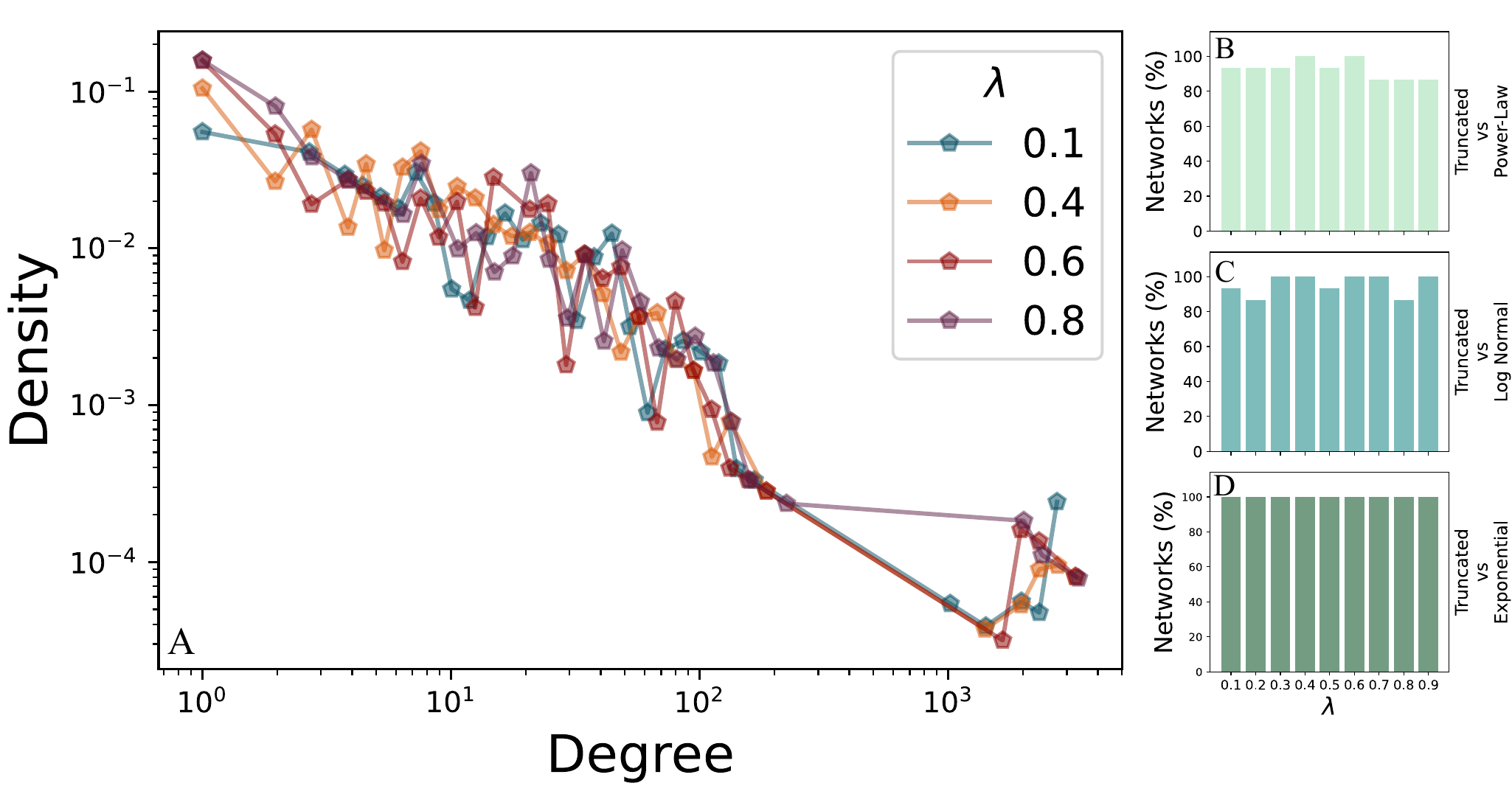}
\caption{\textbf{Degree distribution for the BC-SBM}
Examining the established degree distributions for the networks generated by the bursty-corrected SBM generative model. 
        \textbf{a.} Long-tailed degree distributions for networks generated by the BC-SBM, for different $\lambda$ values.
        \textbf{b., c., d.} The percentage of simulated networks whose degree distributions are better modeled by a truncated power-law model compared to a power-law model, a log-normal model and an exponential model, correspondingly, as determined by a positive log-likelihood ratio and a p-value $<0.1$.
        }\label{fig:simulation_TPL_degree}
        \vspace{-0.4cm}
\end{figure}

We suggested a new Bursty-Corrected Stochastic Blockmodel to enhance the understanding of the forces governing coordination between different entities. 
After demonstrating that the networks generated by our model reproduce the collaborative burstiness seen in empirical experiments, we wish to ascertain that our model also reproduces basic network characteristics, such as a power law degree distribution, regardless of the chosen $\lambda$ parameter, controlling the percentage of intra-community edges \cite{karrer2011stochastic}.
Panel A in Fig. \ref{fig:simulation_TPL_degree} presents the incoming degree distributions of arbitrary simulated networks, for different $\lambda$ values, suggesting that the degree distribution is heavy-tailed. 
Consequently, we calculate the degree distributions for each of the $150$ simulated networks. 
We then apply the statistical analysis methods first presented in \cite{clauset2009power} to evaluate which of four potential heavy-tailed models best fits their degree distributions.
Consistently with the minimal analyzed set requirements presented in \cite{clauset2009power} to guarantee reliable estimates, we consider only networks that have at least $50$ unique degrees. 
For all eligible networks, we perform a goodness of fit test calculating the Log-Likelihood Ratio (LLR) of the different models and the corresponding p-values. 
Panels B, C and D in Fig. \ref{fig:simulation_TPL_degree} present the comparison of the truncated power-law to the power-law, log-normal and exponential models for incoming degree, respectively, for all $\lambda$ values, demonstrating the percentage of networks obtaining a positive LLR and a p-value $< 0.1$. 
These results indicate that across all $\lambda$ values, the truncated power-law model provides a better fit for the majority of simulated networks, compared to all other heavy-tailed models, with high statistical confidence.

Next, we wish to asses the coordination detection performance of different temporal and structural state-of-the-art model, applied over the simulated networks.
Table \ref{tab:simulation_sota_nmi} presents the average NMI of the different models, as a function of the $\lambda$ associated with the simulated network. 
The results manifest a clear advantage of the bursty model in detecting coordination in the simulated data, up to a level of $\lambda=0.7$.
Above this level of intra-community edges, structural models start outperforming the bursty model, illustrating their clear dependency on dense community structure. 

\begin{table}[htbp]\centering
\caption{NMI of coordination detection on simulated networks}\label{tab:simulation_sota_nmi}
\scriptsize
\begin{tabular}{l||c|cccc|cccccc}\toprule
& &\multicolumn{4}{c}{Temporal} &\multicolumn{5}{c}{Structural} \\\midrule
\textbf{} &\textbf{Bursty} &\textbf{pTST} &\textbf{Mmnt.} &\textbf{Chron.} &\textbf{GPT} &\textbf{DW} &\textbf{N2V} &\textbf{LPA} &\textbf{Louv.} &\textbf{Edges} \\
\textbf{\textbf{$\lambda$}} & &\textbf{} &\textbf{} &\textbf{} &\textbf{} &\textbf{} &\textbf{} &\textbf{} &\textbf{} &\textbf{} 
\\\midrule
\textbf{0} &\textbf{0.55} &0.16 &0.19 &0.35 &0.42 &0.48 &5e-16 &5e-16 &1e-7 &5e-16 \\
\textbf{0.1} &\textbf{0.60} &0.19 &0.20 &0.36 &0.39 &0.20 &3e-1 &2e-15 &2e-2 &2e-15 \\
\textbf{0.2} &\textbf{0.57} &0.07 &0.23 &0.40 &0.46 &0.39 &0.49 &5e-16 &8e-3 &5e-16 \\
\textbf{0.3} &\textbf{0.61} &0.07 &0.22 &0.39 &0.42 &0.37 &0.4 &8e-16 &0.01 &0.16 \\
\textbf{0.4} &\textbf{0.64} &0.04 &0.25 &0.41 &0.42 &0.36 &0.47 &2e-16 &8e-3 &0.2 \\
\textbf{0.5} &\textbf{0.73} &0.09 &0.25 &0.40 &0.44 &0.39 &0.43 &2e-16 &5e-3 &0.37 \\
\textbf{0.6} &\textbf{0.76} &0.06 &0.26 &0.44 &0.45 &0.45 &0.49 &5e-16 &5e-3 &0.52 \\
\textbf{0.7} &\textbf{0.74} &0.29 &0.27 &0.46 &0.46&0.35&0.49 &5e-16 &0.02 &0.68 \\
\textbf{0.8} &0.76 &0.29 &0.26 &0.46 &0.47 &0.26 &0.52 &5e-16 &0.03 &\textbf{0.90} \\
\textbf{0.9} &0.74 &0.43 &0.31 &0.46 &0.46 &0.04 &0.19 &5e-16 &0.06 &\textbf{0.94} \\
\textbf{1} &0.72 &0.36 &0.30 &0.44 &0.45 &0.84 &0.9 &\textbf{1.0} &0.96 &\textbf{1.0} \\
\bottomrule
\end{tabular}
\end{table}

 \begin{table}[htbp]\centering
\caption{Communities within a single financial market}\label{si:single_dom_comms}
\scriptsize
\begin{tabular}{l||c|c|cc}\toprule
Group Ind. & $\#$ Entities & $\#$ Trans. \\\midrule
G1 &3  &406\\
G2 &8  &360\\
G3 &4  &9597\\
G4 &4 &336\\
G5 &9  &199133\\
G6 &4  &61066\\
G7 &17  &5261\\
G8 &6  &3509\\
G9 &3  &276\\
\bottomrule
\end{tabular}
\end{table}

\begin{table}[htbp]\centering
\caption{Telegram Communities}\label{si:single_dom_comms_telegram}
\scriptsize
\begin{tabular}{l||c|c|cc}\toprule
Group Ind. & $\#$ Entities & $\#$ Trans. \\\midrule
G1 &1542  &219861\\
G2 &17  &10303\\
G3 &29  &28413\\
G4 &12821 &119910\\
\bottomrule
\end{tabular}
\end{table}

\begin{table}[htbp]\centering
\caption{Communities across financial markets}\label{si:multi_dom_comms}
\scriptsize
\begin{tabular}{l||c|c|cc}\toprule
Group Ind. & $\#$ Entities &  $\#$ Domains & $\#$ Trans. \\\midrule
G1 &13 &15 &63635\\
G2 &7 &11 &92985\\
G3 &20 &14 &32805\\
G4 &5 &5 &1904\\
G5 &5 &12 &9406\\
G6 &3 &3 &818\\
G7 &8 &7 &38076\\
\bottomrule
\end{tabular}
\end{table}
\begin{table}[h]\centering
\caption{Communities across social media platforms}\label{si:SM_cd_comm}
\scriptsize
\begin{tabular}{lrrrr}\toprule
Group Ind. &$\#$ Entities &$\#$ Domains &$\#$ Trans. \\\midrule
G0 &3 &2 &10708 \\
G1 &3 &2 &26083 \\
G2 &3 &2 &5873 \\
G3 &3 &3 &47586 \\
G4 &4 &2 &30852 \\
G5 &3 &2 &1729 \\
G6 &4 &3 &19932 \\
G7 &9 &2 &42449 \\
G8 &3 &3 &29882 \\
G9 &3 &2 &14008 \\
G10 &3 &2 &15511 \\
G11 &3 &2 &7580 \\
G12 &4 &2 &10611 \\
G13 &3 &2 &8573 \\
G14 &3 &2 &42806 \\
G15 &3 &2 &1270 \\
G16 &5 &2 &47212 \\
G17 &4 &3 &11660 \\
G18 &2 &2 &3460 \\
G19 &5 &2 &22736 \\
G20 &3 &2 &4428 \\
G21 &8 &2 &72604 \\
G22 &6 &3 &54607 \\
G23 &9 &2 &58706 \\
G24 &3 &2 &5546 \\
G25 &3 &2 &4331 \\
G26 &3 &2 &472 \\
G27 &3 &2 &12034 \\
G28 &4 &2 &55462 \\
G29 &3 &2 &13222 \\
G30 &3 &2 &3836 \\
G31 &3 &2 &14323 \\
\bottomrule
\end{tabular}
\end{table}

\newpage

\newpage

\newpage

\newpage
\paragraph{Funding Acknowledgment}
Research was sponsored by the United States Air Force Research Laboratory and the Department of the Air Force Artificial Intelligence Accelerator and was accomplished under Cooperative Agreement Number FA8750- 19-2-1000. The views and conclusions contained in this document are those of the authors and should not be interpreted as representing the official policies, either expressed or implied, of the Department of the Air Force or the U.S. Government. The U.S. Government is authorized to reproduce and distribute reprints for Government purposes notwithstanding any copyright notation herein.

\bibliography{community_bib}

\end{document}